\documentstyle[multicol,pre,aps,epsf]{revtex}
\begin{document}

\title{Novel glassy behavior in a ferromagnetic p-spin model}

\author{Michael R. Swift$^{1,3,}$\cite{a1}} 
\address{$^1$School of Physics and Astronomy, University of
Nottingham, Nottingham NG7 2RD, UK} 
\author{Hemant Bokil$^{2,3,}$\cite{a2}}
\address{$^2$Abdus Salam ICTP, Strada Costiera 11, Trieste 34100, Italy}
\author{Rui D.M. Travasso$^3$ and Alan J. Bray$^3$} 
\address{$^3$Department of Theoretical Physics, University of
Manchester, Manchester M13 9PL, UK} 

\date{\today} 
\maketitle

\begin{abstract}
Recent work has suggested the existence
of glassy behavior in a ferromagnetic 
model with a four-spin interaction. Motivated by
these findings, we have studied the dynamics of
this model using Monte Carlo simulations with particular
attention being paid to two-time quantities.
We find that the system shares many features in common with
glass forming liquids. In particular, 
the model exhibits: (i) a very long-lived metastable state,
(ii) autocorrelation functions that
show stretched exponential relaxation, (iii) a non-equilibrium
timescale that
appears to diverge at a well defined
temperature, and (iv) low temperature
aging behaviour characteristic
of glasses. 

\end{abstract}


\begin{multicols}{2}

\section{Introduction}
Many liquids when cooled below the melting temperature 
do not crystallize. Instead, 
as the temperature is lowered, the relaxation time
(or the viscosity) increases dramatically, eventually becoming
so large that the liquid appears frozen on experimental
timescales. For all practical purposes the system has solidified
yet there is no long range order. The system is said to have
become a glass. Despite over fifty years of work, a consensus is lacking on
a theory of this phenomenon\cite{deb}. 
Even as basic a question as the existence of a genuine thermodynamic 
glass transition is difficult to decide experimentally. 
Therefore, much work in this field has been numerical. Simulations
in the continuum of glassy systems remain difficult, and,
despite the substantial progress in this direction\cite{kob},
there are still many unresolved issues.

Because of these problems it would be useful to have simplified lattice
models that behave in some respects like glasses and which do not 
have quenched disorder. Recently,
Lipowski\cite{lip1} and Lipowski and Johnston\cite{lip2,lip3}
studied such a model by Monte Carlo simulations and provided
evidence of glassy behavior. 
The model is defined by the Hamiltonian 
\begin{equation}
H= -J\sum_{\Box} S_i S_j S_k S_l,
\label{eqn1}
\end{equation}
where the spins sit on a cubic lattice and the interactions
are between the four spins at the corners of each plaquette.
The spin variables take values $S_i = \pm 1$ and
$J$ is the strength of the ferromagnetic coupling.
(We will take $J=1$ in all that follows.) 
Their study was motivated
by the fact that the spin-glass version of 
this model was found numerically
\cite{silvio} to have several features in common with structural glasses.
In addition, there is a close connection\cite{lip1} between the plaquette
model, Eq.~\ref{eqn1}, and the Ising model with competing interactions
studied in the context of glasses by Shore and Sethna\cite{ss}. 
Lipowski and Johnston\cite{lip2} showed that, in equilibrium, the
plaquette model has
a first order transition 
as the temperature is lowered but that
this transition is never observed in simulations within accessible timescales.
Instead, one sees what appears to be a
dynamical transition to a glassy state\cite{lip2,lip3}.

The work of these authors concentrated on single time
quantities. To see clear signatures of glassy 
behaviour and to understand which particular aspects of
glasses are present in this model, it is important
to study two time quantities such as autocorrelation
and response functions. We do just that in this paper and
our findings are organized as follows: in section
II, we summarize the results of Lipowski and Johnston 
on the statics of the plaquette model and the
strong metastability that they observed. In section III, we examine
the metastable state in greater detail. In particular, we
calculate the timescale for nucleation and show that
it is indeed extremely large. 
In section IV, we present results for the energy
auto-correlation function in the supercooled
phase which is unambiguously described by a stretched exponential.
We investigate the behavior of the associated timescale as
a function of temperature, showing that it diverges with
a power law at a temperature indistinguishable from the 
lower limit of metastability. This temperature is that which
the authors of \cite{lip1,lip2} called
the glass transition temperature $T_g$. 
In section V, we study the dynamical behavior
below $T_g$. The spin-spin autocorrelation function and overlap distribution
function exhibit aging of a form that is characteristic of 
glasses, showing that the plaquette model has a complex free-energy
landscape. 
These findings provide further evidence to support the conjecture
that the plaquette model exhibits an ideal glass transition.

\section{The Plaquette Model}

In equilibrium, the model defined by Eq.~\ref{eqn1} has a first order
transition at $T_c\simeq 3.6\cite{lip2,esp,beppe}$.
This system possesses a novel symmetry: flipping
all the spins in any plane of the cubic lattice leaves
the Hamiltonian invariant, leading to a degeneracy of ground
states $\sim 2^{dL}$ for a cubic system of size $L$
in $d$ dimensions.
Note, however, that the ground state entropy is finite.
Because of this symmetry, the magnetisation 
is not a good order parameter.
Instead, one can use the internal energy per
spin, $E$, to investigate this model further.

Lipowski and Johnston looked at the dynamics of $E$ following a quench
from a high temperature state to a temperature $T<T_c$\cite{lip2}.
They found different behaviour depending on the value of
$T$. In particular, for $3.4<T<T_c$, $E$ appears
not to relax to its value in the low temperature phase.
Rather it settles into a plateau at an energy 
higher than that of the true low temperature phase at the
same temperature. 
Furthermore, the time the system spends on the plateau (in a ``supercooled''
state) appears to increase at least exponentially with system size,
suggesting the existence of a genuine metastable phase.
Such behaviour , if present, would indeed be strange because
one does not expect true
metastability (i.e. two local minima of the free energy
separated by diverging free energy barriers)
in systems with finite range interactions. Moreover,
nucleation theory tells us that the larger the
system, the shorter the lifetime of a metastable 
state, provided that
the system is large enough to support a nucleating droplet.
To understand whether the strong metastability
is indeed a finite size artifact, Lipowski and Johnston studied
a quench starting from a configuration with a droplet of the low
temperature phase in a high temperature background.
They reported that for sufficiently large seeds, the
system always relaxed to the low temperature phase,
concluding that ordinary nucleation was operative. However, this
does not explain why the supercooled phase appears
so stable on observable timescales. The 
analysis we report in the next section provides an explanation
for this phenomenon.

For quenches to below $T=3.4$ the behaviour was found to be 
quite different. The
plateau disappeared and, after initial transients, 
the energy decreased extremely slowly. Moreover,
it appeared to relax not to the value in the low temperature
phase, but to a value extensively higher.
They identified this state as 
the glassy phase. Finally, they have recently shown that a characteristic
zero-temperature length increases very slowly with the inverse
cooling rate, as is seen in ordinary glasses\cite{lip3}.

To summarize, this model has many
properties that are similar to those seen in real glasses:
a first order transition analogous to crystal melting, 
strong metastability effects in the supercooled phase,
and a kinetic transition to a glassy state characterised 
by slow dynamics.

In the next section we will discuss the
anomalous metastability described above before
looking in some detail at the 
dynamical behavior of this model by studying
the autocorrelation functions both in the supercooled phase
and in the ``glassy'' state. 

\section{Metastability}

We begin by summarizing the basics of nucleation theory\cite{nuc}.
The idea is that if the system is in a metastable
state, the process of finding the true free-energy minimum involves the
formation of a droplet of the stable phase which can
grow and take over the whole system. Phenomenologically,
one can express the free energy barrier to forming a droplet of linear
dimension
$R$ as $\Delta = A\sigma R^{d-1} - B\delta f R^d$ in $d$ dimensions.
Here $\sigma$ is the surface tension between the metastable
and stable phases and $\delta f$ is the difference in
the bulk free energy densities of the two phases. A and B are
constants characterizing the geometrical shape of the droplet. 
Because
the surface term is a cost and the bulk term
is a gain, one can maximize $\Delta$ to find a critical radius
$R^{*} = (A(d-1)/Bd)(\sigma/\delta f)$. Droplets of a radius
greater than $R^{*}$ can grow and take over the whole
systems while those of radius less that $R^{*}$ will
shrink. At low temperatures and 
in three dimensions, the timescale for nucleation of 
a droplet is then given by $\tau_{\rm nuc} \sim 
\exp(4 A^3 \sigma^3/27 B^2 \delta f^2 T)$.

\begin{figure}
\narrowtext
\centerline{\epsfxsize\columnwidth\epsfbox{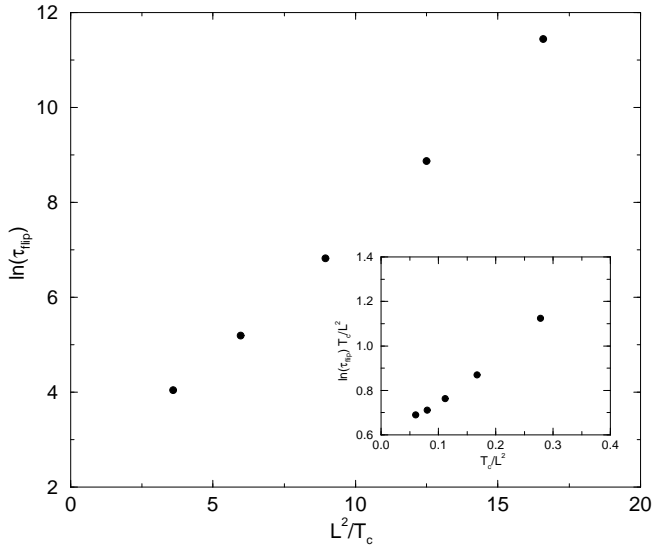}}
\caption{Logarithm of the flip time between high and low 
temperature phases at $T_c$ as a function of system size. 
The errors are smaller than the symbols plotted. The inset shows the
extrapolation for large $L$.} 
\label{Fig1}
\end{figure}

For the plaquette problem, the free energy difference between the two 
phases can be estimated from the data in \cite{lip2}. We find 
$\delta f \approx 0.5 |T-T_c|$. The surface tension $\sigma$ at $T_c$
can be calculated from the size dependence of the mean time 
it takes the system to flip between the high and low temperature
phases. This time $\tau_{\rm flip}$ is given by 
$\tau_{\rm flip} \sim e^{{2\sigma L^2}/ {T_c}}$, where $L$ is the system
size and the factor of $2$ comes from using period boundary conditions.
We have performed standard single spin-flip Monte Carlo simulations
to determine this transition time numerically for systems of size 
$L=4,5,6,7,8$. The transitions are identified from jumps in the energy 
between the values appropriate to the high and low temperature phases. 
For each system size, an `effective transition temperature', $T_c(L)$, is 
determined by the requirement that the system spend equal time in both 
phases, i.e.\ at each of the two possible values of the energy. 
This gives $T_c(L) \simeq 4.45$, 4.19, 4.03, 3.92 and 3.86 for 
$L=4$, 5, 6, 7 and 8 respectively. The mean transition time, 
$\tau_{\rm flip}(L)$, is measured at $T_c(L)$, and the surface tension, 
$\sigma$, obtained by plotting $\ln \tau_{\rm flip}(L)$ against 
$L^2/T_c(L)$. The data is presented in Fig.~\ref{Fig1}. For large $L$, 
the points should approach a straight line with slope $2\sigma$. Since 
there is clear curvature in the data for the small values of $L$ 
available, we attempt an extrapolation to large $L$ by plotting 
$[T_c(L)/L^2]\ln \tau_{\rm flip}$ against $T_c{L}/L^2$, as shown in the 
inset to Fig.~\ref{Fig1}. The intercept on the vertical axis gives $2\sigma$. 
While the extrapolation to $L=\infty$ is necessarily subjective, from 
the curvature of the data it is clear that any reasonable procedure 
will give a value of $2\sigma$ greater than the value $\simeq 0.62$ 
obtained from a linear extrapolation of the last two points, and less than 
the value $\simeq 0.69$ given by the $L=8$ point. To obtain lower bounds 
on critical drop sizes and nucleation times we use the estimate  
$\sigma \simeq 0.31$ obtained from the linear extrapolation. 
Assuming that $\sigma$ depends only weakly on $T$ close to $T_c$, and 
taking $T_c=3.6$, we estimate the critical droplet 
radius at $T=3.5$ to be $R^{*} \approx 25$ for a cubic droplet 
($A=6$, $B=1$) and $\approx 12$ for a spherical droplet 
($A=4\pi$, $B=4\pi/3$). The corresponding time to nucleate
a cubic droplet is of the order of $10^{47}$ MCS (and $\sim 10^{25}$ 
MCS for a spherical droplet). 

These findings provide an explanation for the 
results of \cite{lip2},  in which it was noted that the relaxation time
of the metastable phase grows rapidly with system size. 
In \cite{lip2}, the system sizes simulated were all less than, 
or of the same order as, $R^{*}$. 
In this limit, the free energy barrier to nucleation is set by the 
system size alone, as a droplet of the equilibrium phase cannot fully form.
Consequently,the lifetime of the metastable state will
grow exponentially with the size of the system\cite{lip2} as reported.
We thus conclude that the supercooled state is indeed metastable
but its lifetime is many orders of magnitude larger than
the times accessible in Monte Carlo simulations.

\section{Dynamics in the supercooled phase}

The philosophy underlying the search for models without
disorder that show glassy behavior is that in glasses,
inhomogeneities are self-generated, even though the microscopic Hamiltonian
has no intrinsic disorder. Thus there is
a spontaneous 
separation of the degrees of freedom into variables that exhibit  
either slow or fast relaxation. 
One expects that it is the slow degrees of
freedom that freeze as the glass transition is approached.
However, this separation 
is not necessarily a priori obvious, and one has to 
be careful about which quantities one studies. It might happen,
for example, that the spin-spin autocorrelations 
do not show any anomalous behavior, but that some complicated
functions of the spins show a dramatic slowing down. For this
reason it is often useful to look at autocorrelation functions
of global quantities because these presumably pick out the
longest timescale or the slowest degree of freedom. 

For the system defined by Eq.~\ref{eqn1},
one such quantity is the energy per spin, $E$, and we now discuss
our findings for the energy autocorrelation function
in the range of temperatures $T_g<T<T_c$.
Our results have been obtained from Monte Carlo simulations
using single spin-flip Glauber dynamics. We have considered
cubic systems of linear size $L$, with $L$ ranging from $12$ to $64$.
Most of the data presented here is for a system size of $L=48$.
To measure the energy autocorrelation function at some temperature $T$
we start from a random configuration (infinite temperature)
and quench to the temperature $T$. We then wait for a time $t_w$
and measure the autocorrelation function $A(t,t_w)$ 
for subsequent times, $t$, where
$A(t,t_w) = <E(t_w) E(t_w+t)>$. The angular brackets denote
averages over random initial conditions.
In equilibrium one expects 
$A(t,t_w)$ to be independent of $t_w$ and it is only in equilibrium
that one can define a meaningful timescale associated with relaxation. In
the supercooled phase, we found that equilibrium is easily
achieved in terms of this definition even though the system 
is in a metastable state. In particular, for any
temperature $T_g<T<T_c$, the autocorrelation function becomes
independent of $t_w$ once $t_w$ becomes of the order of $10^3$ MCS. Note
that this time is at least $10$ orders of magnitude smaller than the lifetime
of this metastable state.

\begin{figure}
\centerline{
\epsfxsize=0.9\hsize{\epsfbox{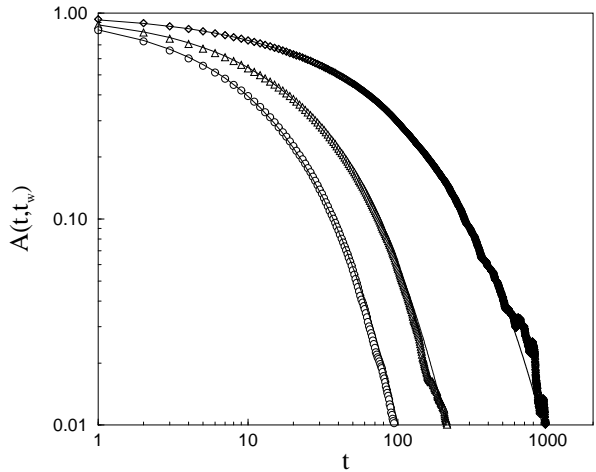}}}
\narrowtext{\caption{The energy autocorrelation
function $A(t,t_w)$ in the supercooled phase. Three temperatures
are shown: $T=3.6 (\circ), 3.5 (\triangle), 3.42 (\diamond)$. The lines
are fits to stretched exponentials, Eq.~\ref{eqn2}, with
parameters: $\tau=11, \beta=0.7$ at $T=3.6$,
$\tau=21, \beta=0.66$ at $T=3.5$ and $\tau=72, \beta=0.6$ at $T=3.42$.}
\label{Fig2}}
\end{figure}

\begin{figure}
\centerline{
\epsfysize=0.8\columnwidth{\epsfbox{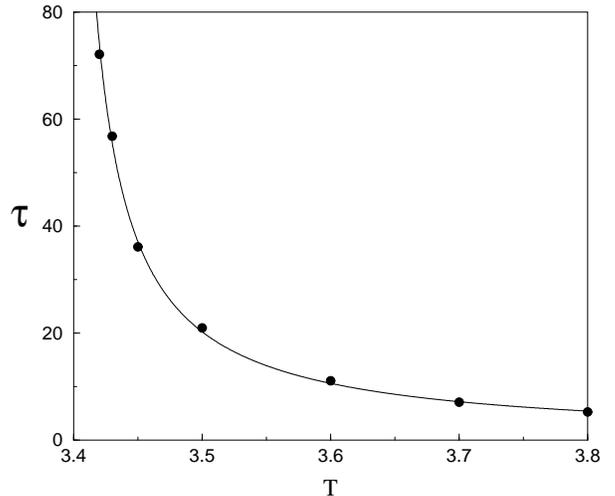}}}
\narrowtext{\caption{The timescale of relaxation $\tau$ as a function
of temperature. The curve is the best fit to a power-law
divergence given by Eq.~\ref{eqn3}.}\label{Fig3}}
\end{figure}

In Fig.~\ref{Fig2} we show $A(t,t_w)$ for three different temperatures
below $T_c$ and above $T_g$, together with attempted fits
to stretched exponentials of the form
\begin{equation}
A(t,t_w)= A_0\, e^{-\left(\frac{t}{\tau}\right)^\beta}.
\label{eqn2}
\end{equation}
As can be seen from the figure, the fits are extremely good and
we can identify a timescale for relaxation, $\tau$, at each temperature.
The exponent $\beta$ is also temperature dependent and is found
to lie in the range $0.5 < \beta < 1$.
In Fig.~\ref{Fig3} we plot the behaviour of $\tau$
as a function of $T$. It appears to increase sharply as $T$ is reduced 
and can be fit quite accurately by a power-law divergence ,
\begin{equation}
\tau=\frac{2.23}{T-3.39},
\label{eqn3}
\end{equation}
as shown in the figure.
This gives a ``critical'' temperature very close to that which was identified
in \cite{lip2} as the lower limit of stability, $T_g$, and
interpreted as the glass transition temperature. 
Both stretched exponential relaxation functions and a diverging
relaxation time are features of real glasses. We also tried a Vogel-Fulcher
(stretched-exponential) fit 
to $\tau$ but found that the corresponding ``critical
temperature'' is significantly lower than $T_g$. As we
will show in the next section, $T_g$ signifies the onset of aging phenomena
and it is thus impossible to define a timescale below this temperature.
(We have also measured the spin-spin autocorrelation
function but found that the behaviour was much harder to interpret.
In fact, the difficulty of interpreting relaxational behavior
of local quantities has been noted before in molecular
dynamics studies of Lennard-Jones systems in two dimensions
\cite{klein}.)
A possible explanation for this divergence is to
note that in mean-field
theory, the plaquette model has a first order transition at 
a temperature quite close to $T_c$\cite{beppe}. In any mean-field
first order transition, one expects a region of metastability
delimited by the spinodal temperature. Above this temperature,
there is genuine metastability and one expects timescales of
autocorrelation functions to diverge as the spinodal
temperature is approached
When one moves away from mean-field,
genuine metastability is no longer present and instead 
the spinodal gets rounded out. Our
results for the autocorrelation functions, and those of the authors
of \cite{lip2} for the energy relaxation, suggest that
the spinodal is actually remarkably well-defined in this
short-range system. It is therefore plausible that what we are calling 
the glass transition
temperature is really a remnant of the mean-field spinodal. This too is
something that is seen in Lennard-Jones systems in two dimensions
\cite{klein}. However, the stretched exponential form of 
$A(t, t_w)$ would not be expected in any simple mean-field treatment.

Thus, the short-range plaquette 
model has a first order transition analogous to crystal melting, but
because of the extraordinarily long-lived supercooled phase, it also
exhibits timescales for relaxation that diverge at $T_g$. 
We would like to emphasize that in this model
$T_g$ is not strictly well-defined. In fact, we expect that as one gets
closer and closer to $T_g$, or waits for longer in the metastable
state, the divergence we are seeing will
get rounded out. Numerically, however, it is not possible to see this
rounding in any reasonable amount of time.
This again is a feature in common with glass forming liquids.

\section{Dynamics in the glassy phase}

\begin{figure}
\centerline{
\epsfysize=0.8\columnwidth{\epsfbox{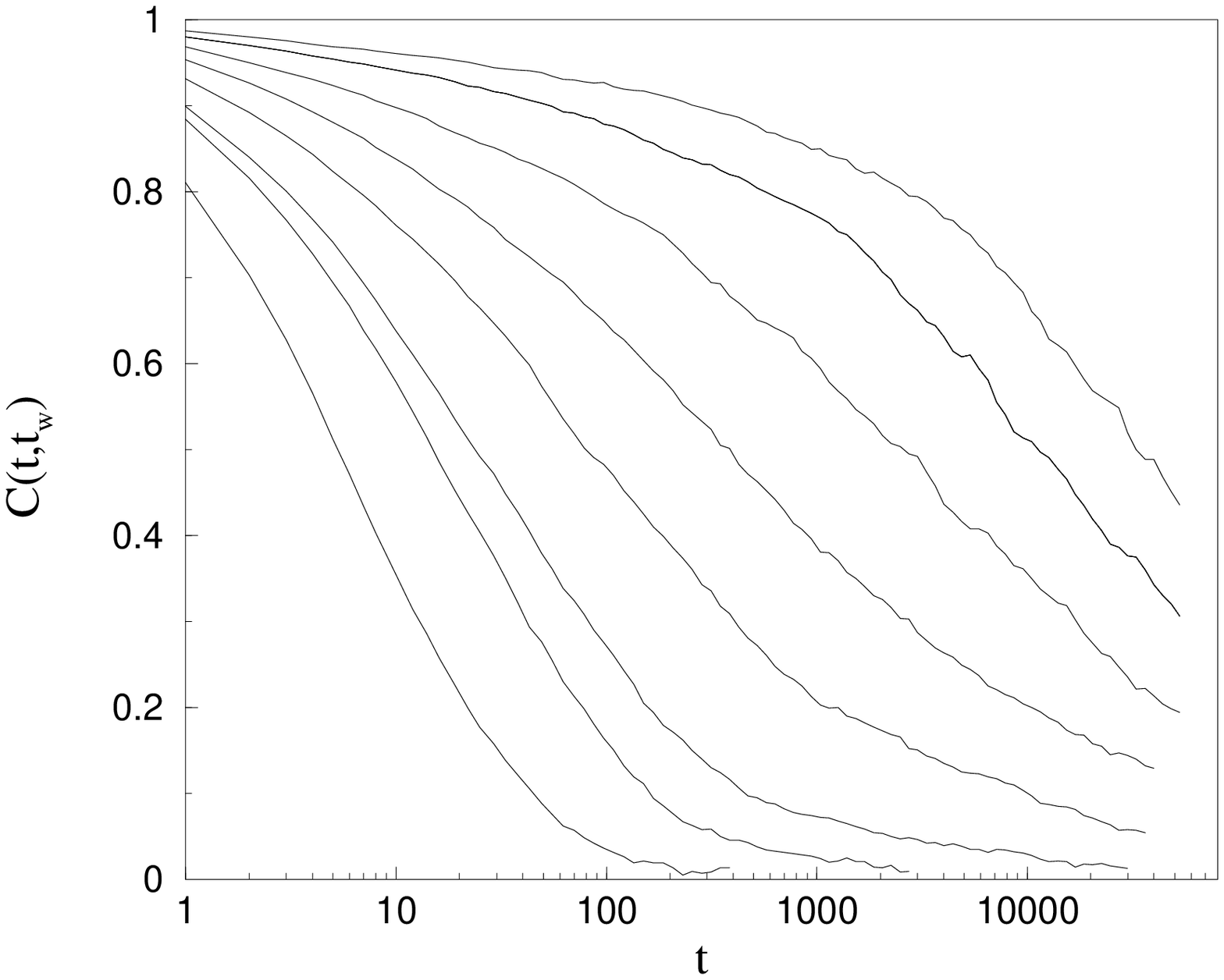}}}
\narrowtext{\caption{The spin autocorrelation
function $C(t,t_w)$ in the glassy phase at temperature $T=3.2$.
Eight different different waiting
times are shown. From bottom to top $t_w= 10$, $100$, $200$, $500$, $1000$,
$2000$, $5000$, $10000$.}
\label{Fig4}}
\end{figure}

For a quench to a temperature $T>T_g$, as we have seen above, the
system becomes trapped in a metastable state for times up to $10^{24}$
MCS. Within this time window, two-time correlation functions
become waiting-time independent and the system behaves as if
in equilibrium. This is no longer the case if $T<T_g$. The system
now continues to evolve slowly for all accessible times and two-time
quantities exhibit aging. To investigate this behaviour we have
measured the spin-spin autocorrelation function,
\begin{equation}
C(t,t_w)=\frac{1}{N} \sum_i S_i(t_w) \, S_i(t+t_w), 
\label{eqn4}
\end{equation}
for a range of waiting-times $t_w$.
Our results are shown in Fig.~\ref{Fig4}. There is a strong dependence
on $t_w$ for all times and temperatures we have considered.
In the long time regime, $C(t, t_w)$ decays as a power of $t$
with an exponent of around $0.35$.
Similar behaviour has been noted in MD simulations of
binary Lennard-Jones mixtures\cite{kob2}.
We have also attempted to collapse the data at a fixed temperature
with a scaling form
$C(t,t_w)=\tilde C(t/\tau(t_w))$ where $\tau(t_w) \sim t_w^{\alpha}$.
While the fit is not perfect, our best estimate for $\alpha$ is
$\alpha=2$. This type of behaviour has been termed super-aging\cite{bouchaud}
as the characteristic timescale for relaxation, $\tau(t_w)$, grows
faster than the weighting time itself.
Note that the spin-glass version of this model exhibits sub-aging with 
$\alpha \approx 0.77$\cite{silvio}.

However, simply having a strong waiting-time dependence in a
two-time correlation function does not imply glassy behaviour.
Indeed, simple coarsening systems exhibit aging\cite{alan}. To distinguish
this type of dynamics, type I aging\cite{barat} , from glassy dynamics,
type II aging, we have considered the overlap distribution function
\begin{equation}
Q(t_w+t,t_w+t)=\frac{1}{N} \sum_i S_i^{(1)}(t+t_w) \,  S_i^{(2)}(t+t_w). 
\label{eqn5}
\end{equation}
This is measured
by first relaxing the system from a disordered state for a time
$t_w$. Then two copies of the system are made, 
$\{S_i^{(1)}\}$ and $\{S_i^{(2)}\}$,
and each is evolved for a further time $t$ with independent thermal noise.

\begin{figure}
\centerline{
\epsfysize=0.8\columnwidth{\epsfbox{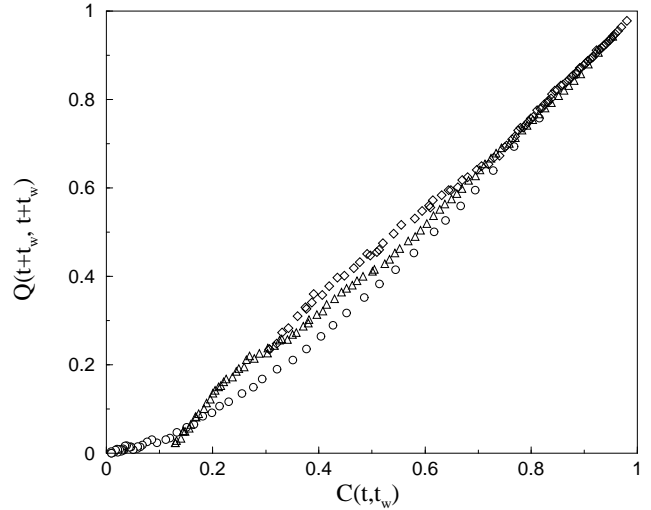}}}
\narrowtext{\caption{Plot of $Q(t_w+t,t_w+t)$ versus $C(t,t_w)$.
Three weighting times are shown: $t_w=100 (\circ), 1000 (\triangle),
5000 (\diamond)$.}
\label{Fig5}}
\end{figure}

In Fig.~\ref{Fig5} we show a plot of $Q(t_w+t,t_w+t)$ vs. $C(t,t_w)$ 
for three different waiting
times. In each case $Q(t_w+t,t_w+t) \to 0$ as $C(t,t_w) \to 0$, indicating
that the two copies of the system continue to move apart from
each other irrespective of when the copies were made. This behaviour
is indicative of type II, glassy aging\cite{barat}.  
Note that
this behaviour is not an automatic consequence of
the existence of infinitely many ground states. For example,
we expect that for a triangular lattice anti-ferromagnet, $Q(t_w+t,t_w+t)$
would tend to a constant as $C(t,t_w)$ tends to zero, despite
the non-zero ground-state entropy per spin.
Finally, we have observed that $C(t, t_w)$ and $Q(t_w+t, t_w+t)$ are related by
$C(2t, t_w)=Q(t_w+t, t_w+t)$ over a wide range of timescales. This
relation is known to hold in equilibrium, where both functions
are independent of $t_w$\cite{barat}. It has also been verified in the aging regime 
of some trap models as well as in the spin-glass version of the
plaquette model\cite{bouchaud}.

These results provide evidence
that the free-energy landscape of the ferromagnetic plaquette model is indeed
complex and has many mutually inaccessible minima. 
This property is similar to what
is seen in structural glasses and in mean field spin-glass models. 

\section{Conclusions}

We have shown that the ferromagnetic plaquette model has many
characteristics in common with glass forming liquids. We
provided an explanation for the long-lived nature of the supercooled
phase in terms of standard nucleation theory. 
By measuring the two-time correlation 
functions, we have observed
two distinct kinds of behavior: (i) in the supercooled phase, the system
appears to be stationary and the timescale of the energy autocorrelation
functions diverges at a temperature identified in 
earlier work as a glass transition temperature $T_g$, and (ii) below
$T_g$ the spin-spin
autocorrelation functions exhibit aging of a form characteristic
of glasses. Our findings, along with the work reported 
in \cite{lip2,lip3}, indicate that a deeper understanding of
this model would indeed be worthwhile.

\medskip

\begin{center}
\begin{small}
{\bf ACKNOWLEDGEMENTS} 
\end{small}
\end{center}

This work was supported by the Engineering and Physical Sciences Research 
Council (MS, HB and AB) and by the Funda\c{c}\~{a}o para a Ci\^{e}cia e a 
Technologia (RT).

\end{multicols} 
\end{document}